\documentclass[twocolumn]{aastex62}

\usepackage{amsfonts}
\usepackage{amsmath}
\usepackage{bm}
\usepackage{graphicx}  
\usepackage{color}
\usepackage{mathrsfs}

\graphicspath{{./}}

\submitjournal{ApJ}

\shorttitle{Kelvin-Helmholtz instability at proton scales with an exact kinetic equilibrium}
\shortauthors{Settino et al.}

\begin{document}

\title {Kelvin-Helmholtz instability at proton scales with an exact kinetic equilibrium}

\correspondingauthor{Adriana Settino}
\email{adriana.settino@unical.it}

\author{A. Settino}
\affiliation{Dipartimento di Fisica, Universit\`a della Calabria, 87036 Rende (CS), Italy}
\author{F. Malara}
\affiliation{Dipartimento di Fisica, Universit\`a della Calabria, 87036 Rende (CS), Italy}
\author{O. Pezzi}
\affiliation{Gran Sasso Science Institute, I-67100 L’Aquila, Italy}
\affiliation{INFN/Laboratori Nazionali del Gran Sasso, I-67100 Assergi (AQ), Italy}
\author{M.Onofri}
\affiliation{TAE Technologies Inc., PO Box 7010, Rancho Santa Margarita, CA 92688, USA}
\author{D. Perrone}
\affiliation{ASI -- Italian Space Agency, via del Politecnico snc, 00133 Rome, Italy}
\author{F. Valentini} 
\affiliation{Dipartimento di Fisica, Universit\`a della Calabria, 87036 Rende (CS), Italy}

\begin{abstract}
The Kelvin-Helmholtz instability is a ubiquitous physical process in ordinary fluids and plasmas, frequently observed also in space environments. In this paper, kinetic effects at proton scales in the nonlinear and turbulent stage of the Kelvin-Helmholtz instability have been studied in magnetized collisionless plasmas by means of Hybrid Vlasov-Maxwell simulations. The main goal of this work is to point out the back reaction on particles triggered by the evolution of such instability, as energy reaches kinetic scales along the turbulent cascade. Interestingly, turbulence is inhibited when Kelvin-Helmholtz instability develops over an initial state which is not an exact equilibrium state. On the other hand, when an initial equilibrium condition is considered, energy can be efficiently transferred towards short scales, reaches the typical proton wavelengths and drives the dynamics of particles. As a consequence of the interaction of particles with the turbulent fluctuating fields, the proton velocity distribution deviates significantly from the local thermodynamic equilibrium, the degree of deviation increasing with the level of turbulence in the system and being located near regions of strong magnetic stresses. These numerical results support recent space observations from the Magnetospheric MultiScale mission of ion kinetic effects driven by the turbulent dynamics at the Earth's magnetosheath (Perri {\it et al.}, 2020, JPlPh, 86, 905860108) and by the Kelvin-Helmholtz instability in the Earth's magnetosphere (Sorriso-Valvo {\it et al.}, 2019, PhRvL, 122, 035102).
\end{abstract}

\section{Introduction} \label{sec:intro}
The Kelvin-Helmholtz instability (KHI) is a phenomenon that can develop in both fluids and plasmas, in configurations where velocity shears are present. During KHI, perturbations are generated in form of a chain of vortices, located along the shear layer, which grow in time starting from infinitesimal fluctuations. In the case of a magnetized plasma, the magnetic field has a stabilizing effect with respect to KHI. Typically, a configuration is unstable when the jump in the bulk velocity across the shear layer is larger than a threshold, which is of the order of the component of the Alfv\'en velocity parallel to the bulk velocity \citep{Chandra1961}. When unstable modes reach a sufficiently large amplitude, they start interacting among them, fragmenting and generating structures at increasingly smaller scales. Moreover, vortices tend to merge forming larger coherent structures, moving part of the fluctuating energy to larger scales. These phenomena lead to a final turbulent state where part of the kinetic energy associated with the velocity shear is dissipated. Therefore, KHI represents a way for a fluid or a plasma to give rise to a turbulent scenario and to convert large-scale motion energy into heat.

KHI has been considered in many natural systems, such as in terrestrial, heliospheric and astrophysical contexts. For instance, (i) KHI has been observed at planetary magnetospheres \citep{Kivelson1995,Seon1995,Fairfield2000,Fairfield2003,Hasegawa2004,Hasegawa2006,Nykyri2006}; (ii) it has been invoked to explain the penetration of solar wind into cometary ionospheres \citep{Ershkovich1983}; (iii) it has been considered in turbulence models at the interface between fast and slow solar wind streams \citep{Roberts1991,Roberts1992}; and (iv) it has been observed in the solar corona at the surface of coronal mass ejections \citep{Foullon2011}. Moreover, the role of KHI has been also studied in the generation of astrophysical jets in relativistic magnetized plasmas \citep{Hamlin2013} or at the interface between the accretion disk and the magnetosphere of a slowly rotating magnetized star \citep{Lovelace2010}, as well as in black holes and neutron stars \citep{Li2004}. KHI is thought to be responsible for the plasma transport across the Earth's magnetopause, during periods of both northward and southward orientation of the interplanetary magnetic field \citep{Foullon2008,Kavosi2015}. 

The unprecedented high-resolution observations conducted by the NASA  Magnetospheric MultiScale (MMS) mission, launched in March 2015, have allowed to inspect KHI onset at kinetic scales \citep{Stawarz2016,Hwang2020}. {\it In-situ} measurements, supported also by numerical simulations, suggest that magnetic reconnection induced by KHI breaks the frozen-in condition, thus favoring the solar-wind plasma entry into the Earth's magnetosphere \citep{Nakamura2017,Eriksson2016,Sisti2019}. Moreover, primary and secondary KHI have been associated to the generation and shaping of flux ropes \citep{Hwang2020,Zhong2018,Zhou2017}. The interconnection between turbulence development and KHI at the non-linear stage has been recently studied by comparing MMS observations with both magnetohydrodynamics (MHD) \citep{Hasegawa2020,Nakamura2020} and hybrid kinetic simulations \citep{Franci2019}. Finally, KHI is supposed to be dawn-dusk asymmetric owing to the different vorticity at the two flanks. This mid-latitude asymmetry has been investigated by means of simultaneous {\it in-situ} observations of THEMIS and MMS satellites \citep{Lu2019}.

KHI in magnetized plasmas has been widely studied in various configurations. Several theoretical studies have been carried out within the MHD framework. The linear stage of the instability, when unstable modes grow exponentially in time, has been investigated for different spatial profiles of the bulk velocity ${\bf u}$ and density, and different orientations of the magnetic field ${\bf B}$ with respect to ${\bf u}$ \citep[see \textit{e.g.}][]{Axford1960,Walker1981,Miura1982,Contin2003}. Moreover, the interplay with tearing instability has been also considered for inhomogeneous magnetic field profiles \citep{Wesson1990}. Dispersive or kinetic effects come into play when the shear layer thickness is of the order of ion length scales (ion inertial length and/or Larmor radius). These phenomena affect the growth rate of unstable modes, which in this regime depends on the relative orientation between magnetic field and vorticity \citep{Nagano1979,Huba1996,Cerri2013}.

The nonlinear evolution of KHI has been numerically studied in a large number of investigations, using both fluids (MHD, Hall-MHD and two-fluid) and kinetic approaches. In fluid simulations, it has been shown that viscosity generates momentum transfer between flows on the two sides of the shear layer \citep{Miura1982}. Moreover, in the case of perpendicular magnetic field, if the simulation box is larger than the vortex length, an inverse cascade takes place where KHI-generated vortices merge forming structures at larger scales (i.e., vortex pairing) \citep{Miura1997,Miura1999a}. This effect has been proposed as a way to follow the time evolution of KHI in non-periodic configurations \citep{Mills2000,Wright2000}, such as at the Earth's magnetopause \citep{Miura1999b}. In the fully nonlinear regime, secondary instabilities can develop, such as Rayleigh-Taylor, secondary KH, or kink-like instabilities, which can compete with the pairing process leading to the disruption of vortices \citep{Matsumoto2004,Nakamura2004,Faganello2008}. Furthermore, in configurations where the in-plane magnetic field component changes sign across the shear layer, magnetic reconnection can couple with KHI, thus creating a magnetic connection between the two sides of the shear layer, with consequences on the transport properties. However, even when the in-plane magnetic component keeps the same sign, reconnection takes place during the nonlinear stage, leading to the formation of complex magnetic topologies (for a detailed discussion see the review by \citet{Faganello2017} and references therein). These phenomena take part to the more general problem of reconnection in small-scale structures generated by turbulence \citep{Servidio2011a,Servidio2011b,Servidio2012}.

In cases when KHI develops in collisionless plasmas at scales of the order of ion scales, such as in the Earth's magnetosphere, kinetic simulations appear to be more suitable than fluid approaches \citep{Pritchett1984,Matsumoto2006,Cowee2009,Matsumoto2010,Nakamura2010,Nakamura2011,Nakamura2013,Henri2013,Karimabadi2013}. The interplay of KHI with other kind of instabilities, such as lower-hybrid drift instability, has been very recently considered by \citet{Dargent2019}. Furthermore, kinetic effects can be important during the nonlinear stage of the instability, when vortices mix and a turbulent state develops. Indeed, kinetic simulations of turbulence at ion scales have highlighted the formation of small-scale structures in the physical space closely related to the generation of out-of-equilibrium features in the particle velocity space, such as temperature anisotropy, agyrotropy of the ion velocity distribution, and/or beams of suprathermal particles \citep{Servidio2012,Greco2012,Perrone2013,Servidio2014,Valentini2014,Servidio2015,Rossi2015,Valentini2016,
Pezzi2017a,Pezzi2017b,Pezzi2017c}. In this perspective, the development of an enstrophy phase-space cascade, due to turbulent fluctuations, has been also proposed in several works \citep{Schekochihin2016,Servidio2017,Eyink2018} and recently observed in the terrestrial magnetosheath \citep{Servidio2017} as well as in kinetic numerical simulations \citep{Pezzi2018,Cerri2018}. Moreover, evidences of the existence of turbulence-driven ion beams in the KHI has been reported in the Earth's magnetosphere~\citep{Sorriso2019,Perri2020}. These phenomena are related to the general problem of understanding cross-scale energy transfer and dissipation in collisionless plasmas \citep{Servidio2015}, such as, for instance, in the solar wind or magnetosphere \citep{Sorriso2018,Sorriso2019}.

Within the kinetic description of KHI, setting up the unperturbed state is a non-trivial issue, that has consequences on the instability onset. Indeed, when a plasma displays inhomogeneities, such as bulk velocity and/or magnetic shears, the simplest way to give a kinetic representation of those configurations is to adopt shifted-Maxwellian (SM) distribution functions (DFs), where parameters like density, bulk velocity and/or temperature vary in space \citep{Umeda2014}. However, in general SMs are not stationary solutions and this could affect the development of the KHI. Typically, this kind of DFs tends to relax generating undamped oscillations with periods of the order of the ion gyroperiod \citep{Nakamura2010,Cerri2013,Henri2013}, which can lead to a modification of the DF, mainly in situations where the vorticity is anti-parallel to the magnetic field. This has an effect on the dispersion relation: for instance, in the perspective of studying the Dawn-Dusk asymmetry of the KHI in the magnetosphere, where the relative vorticity-magnetic field alignment is opposite on the two sides of magnetosphere, using a SM could lead to not completely reliable results. Of course, these phenomena are more relevant when the velocity shear width is of the order of ion scales. 
These effects could be avoided if an exact kinetic stationary DF is employed instead of a SM. Within the framework of fully kinetic theory, this kind of solutions has been proposed in the case of a uniform perpendicular magnetic field \citep{Ganguli1988,Nishikawa1988,Cai1990}, for a nonuniform magnetic configurations \citep{Mahajan2000} and for parallel magnetic field \citep{Roytershteyn2008}, where the SM is enough manageable and easy to be implemented. However, despite of the above-described problems, this kind of solutions has been rarely employed to study the KHI. 

In order to describe phenomena at scales comparable with ion scales, a successful numerical approach is represented by the hybrid Vlasov-Maxwell (HVM) model, where ions are kinetically described by the Vlasov equation, while electrons are treated as a massless fluid \citep{Valentini2007}. In the last decade, this model has been adopted for describing several phenomena occurring at scales where the kinetic ion physics starts to play a significant role into the plasma dynamics \citep{Servidio2012,Matthaeus2014,Franci2015,Servidio2015,Valentini2016,Cerri2016,Cerri2017,Valentini2017}. Within the HVM framework, \citet{Cerri2013} has presented a method to derive approximately stationary ion DFs, based on the evaluation of finite Larmor radius effects in the ion pressure tensor. This approach has been used to describe temperature anisotropy in the presence of shear flows \citep{Cerri2014,DelSarto2016}. However, the solution proposed by \citet{Cerri2013} is not exactly stationary, since small amplitude oscillations develop, even if definitely weaker than those found for a SM. Recently, \citet{Malara2018} have found exact stationary solutions of the HVM equations, describing a magnetized shear flow, in the cases of both parallel and perpendicular uniform magnetic field. These solutions, recently adopted to investigate the production of kinetic Alfv\'en waves in a velocity shear  \citep{Maiorano2020}, differ from the SM close to the shear layer, where temperature anisotropies and agyrotropies are observed in the exact equilibrium. Moreover, in the case of perpendicular magnetic field some moments of the DF are different according to the relative vorticity-magnetic field orientation.

In the present paper we use the HVM model to study the development of the KHI in a configuration with a uniform magnetic field perpendicular to the shear flow. Such a configuration can be representative of the region across the Earth's magnetopause. One of the aims of this study is to establish to what extent adopting the exact stationary solution (EE) instead of the SM distribution function can affect the linear and the nonlinear stages of the KHI. For such a purpose we will compare the time evolution obtained using both the EE found by \citet{Malara2018} and a SM DF, corresponding to the same shear flow. Our results show that using the exact solution affects the values of growth rates, and, to a larger extent, the nonlinear development of the instability, giving origin to a more developed turbulence and larger values for the current density. These results are relevant in the perspective of correctly evaluating the spectral energy transfer and dissipation generated by the KHI.

The plan of the paper is the following: in Section \ref{sec:setup} we describe the initial setup of the simulations with a focus on the equations of the model, the characteristics of the DFs and the perturbations introduced. An insight into the derivation of the EE solution is also provided. 
In Section \ref{sec:numres} we discuss simulations results, directly comparing the EE and SM data. Finally, we give the conclusions in Section \ref{sec:concl}.    

\section{Simulation setup and initial conditions} \label{sec:setup}
To perform the numerical analysis of the KHI, retaining kinetic effects at proton scales, we employed the HVM numerical code \citep{Valentini2007}. The HVM algorithm solves numerically the Vlasov equation for the proton DF, self-consistently coupled to the Maxwell equations for electromagnetic fields, while electrons are treated as a massless fluid. We considered two 
shared-flow initial conditions, with two different initial proton DFs: the exact shared-flow HVM equilibrium distribution, $f_0^{^{(EE)}}$, derived in \citet{Malara2018} and a SM distribution, $f_0^{^{(SM)}}$. Results obtained starting from these two initial conditions will be discussed and compared in Section~\ref{sec:numres}.

The HVM equations are numerically solved in a 2.5D-3V phase-space domain, that is, fully three-dimensional in velocity space while, in physical space, all vectors have three components depending only on two variables $(x,y)$.
Quasi-neutrality condition is assumed and the displacement current is neglected in the Amp\`ere equation, in such a way to discard light waves. 

In dimensionless units, HVM equations are:
\begin{eqnarray}\label{hvm_i}
& &\frac{\partial f}{\partial t}+{\bm v}\cdot {\nabla f}+({\bf E}+{{\bm v}\times {\bf B}})\cdot\nabla_v f=0\\
\label{hvm_m}
& &{\bf E}=-{\bf u}\times{\bf B}+\frac 1 n {\bf j}\times{\bf B}-\frac 1 n \nabla P_e\\
& &\frac{\partial {\bf B}}{\partial t}=-\nabla\times {\bf E}; \;\;\; \nabla\times {\bf B}={\bf j}
\label{hvm_f}
\end{eqnarray}
\noindent 
being $f=f(x,y,v_x,v_y,v_z)$ the proton DF, $\nabla=(\partial_x,\partial_y)$, $\nabla_v=(\partial_{{v}_x},\partial_{{v}_y},\partial_{{v}_z})$, ${\bf E}$ and ${\bf B}$ respectively the electric and magnetic fields, $n$ and ${\bf u}$ respectively the proton density and bulk velocity, computed as the first two velocity moments of $f$, and ${\bf j}$ the total current density. For the electron pressure, we assume an isothermal equation of state, $P_e=nT_{e}$, in the case of SM initial condition, where $n_e=n_p=n$ for the quasi-neutrality assumption and $T_e=\bar{T}$, being $\bar{T}$ the proton temperature far from the shear. On the other hand, for the EE initial condition, as extensively discussed in \citet{Malara2018}, we need to relax the electron closure in order to maintain the equilibrium, by treating the electron pressure, $P_e$, as a further independent quantity determined by the following equation:
\begin{equation}\label{adiab}
\left[\frac{\partial}{\partial t} + ({\bf u}_e\cdot\nabla)\right]\left(\frac {P_e} {n^{\gamma_e}}\right)=0 
\end{equation}
where $\gamma_e =5/3$ is the electron adiabatic index and ${\bf u}_e = {\bf u}-{\bf j}/n$ is the electron bulk velocity.

In Eqs. (\ref{hvm_i})-(\ref{adiab}), time is scaled by the inverse proton cyclotron frequency, $\Omega_{cp}$, velocities by the Alfv\'en speed, $v_A=B_0/\sqrt{4\pi \bar{n} m_p}$ (where $B_0$ is the background magnetic field, $\bar{n}$ the proton density away from the shear regions and $m_p$ the proton mass), lengths by the proton skin depth, $d_p=v_A/\Omega_{cp}$, the magnetic field by $B_0$, the electric field by $v_A B_0/c$, the density by $\bar{n}$, and the electron pressure by $\bar{n} m_p v_A^2$. The development that follows will be expressed in terms of the above dimensionless quantities.

Spatial domain $D=[0,L_x] \times [0,L_y]$ ($L_x=L_y=L=100$) is discretized on a uniformly spaced grid with $N_x=N_y=256$ grid points; periodic boundary conditions have been implemented in the spatial domain. Velocity-space domain is discretized on a uniform grid with $N_{v_j}=71\, (j=x,y,z)$ grid points in each direction. Vanishing boundary conditions have been implemented: $f(|v_j| > v_{max})=0$, being $v_{max}=7 v_{th}$ and $v_{th}=(\bar{T})^{1/2}$ the proton thermal speed; the proton plasma beta is $\beta=2v^2_{th}/v_A^2=2$.

The unperturbed configuration is characterized by: (i) a sheared bulk velocity field ${\bf u}=u(x) {\bf e}_y$, that is directed along $y$ and varies in the $x$ direction; (ii) a perpendicular uniform magnetic field ${\bf B}=B_0 {\bf e}_z$, with $B_0=1$; and (iii) an electric field ${\bf E}=E(x) {\bf e}_x$, whose profile $E(x)$ is related to the bulk velocity ${\bf u}(x)$. In the above expressions ${\bf e}_x$, ${\bf e}_y$ and ${\bf e}_z$ are the unit vectors in the directions of the three Cartesian axes. 

In the case of SM configuration the bulk velocity has the form ${\bf u}=U_y(x){\bf e}_y$, where the function $U_y(x)$ describes the double shear profile:
\begin{equation}\label{Uy}
U_y(x)=U_0\left[\tanh \left( \frac{x-x_1}{\Delta x}\right) - \tanh \left( \frac{x-x_2}{\Delta x}\right) -1 \right]
\end{equation}
Here, $x_1=L_x/4$ and $x_2=3L_x/4$ are the positions of the shears, $\Delta x = 2.5$ is the shear width and $2U_0=2v_A$ is the velocity jump. We point out that the velocity shear has been replicated along the $x$ direction to satisfy periodic boundary conditions. 
The separation between the two shears is not large enough to avoid their interaction during the late nonlinear and turbulent phases of the KHI. Indeed, as reported in Sect.\ref{sec:numres}, vortices of each shear start to attract and merge during the fully-nonlinear phase of the simulation, thus making it impossible to analyze the final stage of the KHI dynamics separately for each shear.

\begin{figure} [!ht]
\centering
\includegraphics[width=0.5\textwidth]{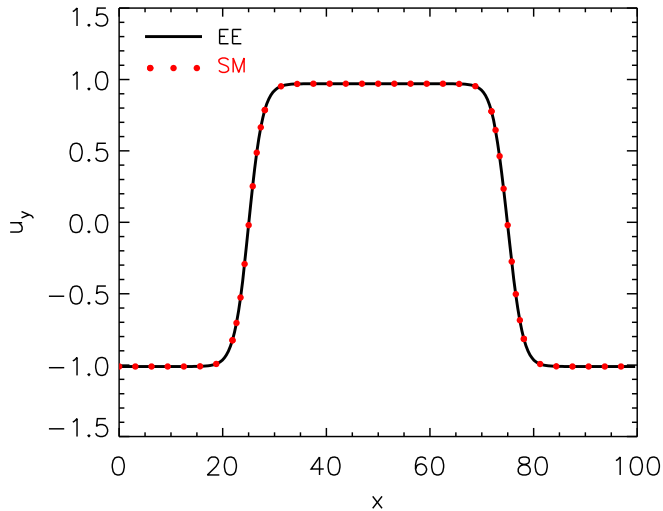}
\caption{The $y$ component of the proton bulk velocity as a function of $x$ for the distribution $f_0^{^{(EE)}}$ (black curve) and $f_0^{^{(SM)}}$ (red dots).}
\label{fig:Eq}
\end{figure}

The profile of the $y$ component of the proton bulk velocity, $u_y$, as a function of $x$ for the SM distribution function is reported as red dots in Fig.~\ref{fig:Eq}, where the presence of the two shear layers is clearly visible.

\subsection{Exact solution}
Beside the SM DF, we considered the stationary solution found by \citet{Malara2018} for the system of HVM equations. In the following, we briefly revisit the derivation and properties of such solution, while more details can be found in \citet{Malara2018}. 
An exact stationary solution of the Vlasov equation, Eq. (\ref{hvm_i}), can be written as a function of constants of single particle motion. Therefore, we consider the motion of a proton in the above electric and magnetic fields. The following relation between the $x$-position and the $y$-component of the particle velocity is found (in dimensionless units):
\begin{equation}\label{x-vy}
v_y(t)= W_0- x
\end{equation}
where $W_0$ is a constant determined by initial conditions. The particle motion in the $x$-direction corresponds to that of a nonlinear oscillator, whose effective potential energy has the form:
\begin{equation}\label{Ueff}
U_{\rm eff}(x) = \Phi(x;x_0) + \frac{1}{2}  \left(x-W_0\right)^2 + \frac{1}{2} v_{0y}^2
\end{equation}
where $\Phi(x;x_0) = -\int_{x_0}^x E(x')dx'$ is the electrostatic potential which vanishes at a given position $x_0$, and $v_{y0}=W_0-x_0$. In Eq. (\ref{Ueff}) energies are normalized to $m_p v_A^2$. Assuming that the bulk velocity profile is uniform away from the shear layers, which corresponds to a uniform electric field, from Eq. (\ref{Ueff}) it follows that the particle motion along $x$ is periodic within a potential well. Therefore, we can define the guiding center position $x_c$ and velocity $v_{yc}$ as the average $x$ position and average $v_y$ velocity, respectively: $x_c=\langle x \rangle_t$, $v_{yc}=\langle v_y \rangle_t=W_0-x_c$. In particular, the point $x_0$ where $\Phi(x;x_0)$ is null is chosen as $x_c$. The reduced energy is defined as:
\begin{equation}\label{redE}
\mathscr{E}_0(x,v_x,v_y,v_z)=\frac{1}{2} \left( v_x^2+v_y^2+v_z^2 \right) + \Phi(x;x_c) - \frac{1}{2}  v_{yc}^2
\end{equation}
The total energy (kinetic + potential) and $v_{yc}$ are both constants of motion. Therefore, $\mathscr{E}_0$ is another constant of motion, equal to the total energy minus the kinetic energy associated with the drift motion. We define a distribution function
\begin{equation}\label{fEE}
f_0^{^{(EE)}}(x,v_x,v_y,v_z)=C \exp \left[ -\frac{\mathscr{E}_0(x,v_x,v_y,v_z)}{v_{th}^2} \right]
\end{equation}
with $C$ and $v_{th}$ constants. Since $f_0^{^{(EE)}}$ is a combination of constants of motions, it is an exact stationary solution of the Vlasov equation. It can be shown that far from the shear layer $f_0^{^{(EE)}}$ reduces to a shifted Maxwellian centered around the drift velocity $(-E/B_0) {\bf e}_y$.
The density $n_0$ associated to $f_0^{^{(EE)}}$ is spatially uniform everywhere except in the regions corresponding to the velocity shears  \citep[see, for details,][]{Malara2018}. 

In the general case, the explicit form of $f_0^{^{(EE)}}$ is numerically calculated on the grid in the 4D phase space $\left\{x,v_x,v_y,v_z\right\}$. For each grid point the particle trajectory is integrated until it closes in the $v_xv_y$ plane, calculating the corresponding values for the constants of motion: the guiding center position $x_c=\langle x\rangle_t$ and velocity $v_{yc}=\langle v_y \rangle_t$; the kinetic energy; and the potential $\Phi(x;x_c)$. Those values are used to calculate $f_0^{^{(EE)}}$ at the given grid point. Results show that the bulk velocity is directed along $y$, \textit{i.e.} ${\bf u}=u(x){\bf e}_y$, and hence the term $-{\bf u} \times {\bf B}$ 
in Eq. (\ref{hvm_m}) is directed along $x$. Choosing a form for the electric field, Eq. (\ref{hvm_m}) is exploited to determine the electron pressure $P_e$. In particular, we adopted the expression $E(x)=-B_0 U_y(x)$, where $U_y(x)$ is the bulk velocity associated with the SM distribution function in Eq. (\ref{Uy}).

In Fig. \ref{fig:Eq} the corresponding profile of $u_y(x)$ from the exact equilibrium solution is plotted (black curve). It can be seen that the bulk velocity profiles corresponding to the exact solution and to the shifted Maxwellian are very close to each other. Nevertheless, we will show that the time evolution of the KHI is different in the two cases. 
Slightly larger differences are found in the density and temperature profiles, which are homogeneous in the case of the shifted Maxwellian, while in the exact equilibrium case they exhibit a maximum and a minimum, localized at the two shears. Percentage variations of these quantities are $17.7\%$ and $12.2\%$, respectively.
Moreover, $f_0^{^{(EE)}}$ exhibits a clear temperature anisotropy in regions close to the shears, being elongated in a direction transverse to the background magnetic field, while reduces to a shifted Maxwellian far from the shears \citep{Malara2018}.

\subsection{Initial perturbation}
At $t=0$, we perturbed the initial configuration through a broadband spectrum of bulk velocity fluctuations. Such perturbations have only $y$ spatial dependence and are generated in the form of random noise. We excited the first $32$ modes in the spectrum with random phases.
For both EE and SM simulations, we summed to the unperturbed function ($f_0^{^{(EE)}}$ or $f_0^{^{(SM)}}$, respectively) the perturbation, shaped as a Maxwellian function shifted in the $v_x$ and $v_y$ directions; that is $f^{^{(EE)}}=f_0^{^{(EE)}}+f_1$ or $f^{^{(SM)}}=f_0^{^{(SM)}}+f_1$, where $f_1$ is defined (in scaled units) as follows:
\begin{multline}
f_1(y,{\bf v})= \frac{n_1}{(\pi\beta)^{3/2}}\exp\left\{-\frac{\left[v_x-u_{1x}(y)\right]^2}{\beta}+\right.\\
-\left.\frac{\left[v_y-u_{1y}(y)\right]^2}{\beta}-\frac{v_z^2}{\beta}\right\};
\label{eq:f1}
\end{multline}
here, $n_1=0.01$ is the amplitude of the perturbation, and $u_{1x}= \sum_{i=1}^{32} cos(k_{y,i} y +\psi_i)$, $u_{1y}=\sum_{i=1}^{32} sin(k_{y,i} y +\phi_i)$, where $k_{y,i}=i2\pi/L$ and $\psi_i$ and $\phi_i$ random phases.
For both the initial perturbed EE and SM distributions, the proton density $n$ and the bulk velocity ${\bf u}$ can be written as:
\begin{equation}
n=n_0+n_1; \;\;{\bf u} =\frac{n_0 {\bf u}_0+ n_1 {\bf u}_1}{n_0+n_1}
\end{equation}
where $n_0$ and ${\bf u}_0$ are the density and the bulk velocity of $f_0$, respectively. For small amplitude perturbations,  the above equation for the bulk velocity can be Taylor expanded in series of $n_1/n_0$ leading to:
\begin{equation}
    {\bf u}={\bf u}_0+\frac{n_1}{n_0}{\bf u}_1 
\end{equation}

Since $n_1$ is constant and $n_0$ is uniform (except in the shear regions for $f_0^{^{(EE)}}$), the perturbed part of the bulk velocity is largely shaped by ${\bf u}_1$, for both EE and SM cases.
We remark that the bulk velocity fluctuations which perturb the initial equilibrium are consistent with field perturbations. Indeed, according to the HVM system of equations (eqs.\ref{hvm_i}-\ref{hvm_f}), the electric field is evaluated by means of  the Ohm's law.
  
\section{Numerical results} \label{sec:numres}
\begin{figure}[ht]
\centering
\begin{minipage}[ht]{\linewidth}
   \centering
\includegraphics[width=\textwidth]{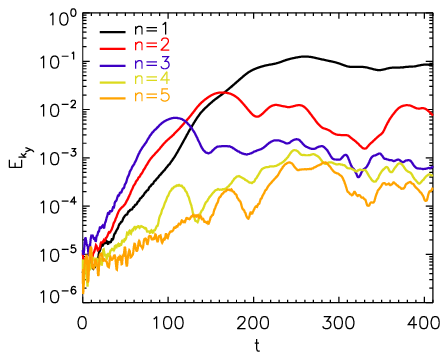}
\end{minipage}
\begin{minipage}[ht]{\linewidth}
   \centering
\includegraphics[width=\textwidth]{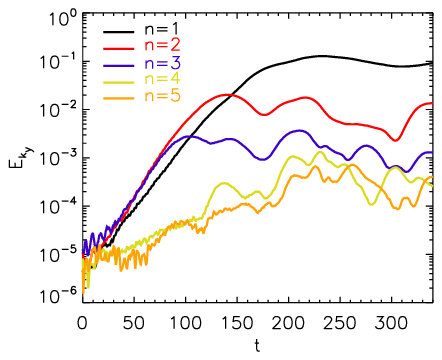}
\end{minipage}
\caption{Time evolution of the first $5$ Fourier components of the spectral kinetic energy $E_{k_y}$ for the EE (top panel) and SM (bottom panel) simulations. }
\label{fig:growth}
\end{figure}

During the linear phase of the instability, the exponential growth of the energy $E_{k_y}$ of the velocity Fourier components perturbed at $t=0$ is observed. In Fig. \ref{fig:growth}, we report the time evolution of $E_{k_y}=\langle|\mathbf{u}_{k_y}(x,t)|^2\rangle_x$, where ${\bf{u}}_{k_y}(x,t)$ is obtained by Fourier transforming ${\bf u}(x,y,t)$ along the $y$ direction and $\langle\cdots \rangle_x$ indicates average over $x\in \left[0,L/2\right )$, for the EE (top panel) and the SM (bottom panel) simulations. After the initial exponential growth, nonlinear saturation is reached for both simulations and no significant differences between EE and SM cases are recovered.
 
\begin{figure*}[ht]
\centering
 \includegraphics[width=\textwidth]{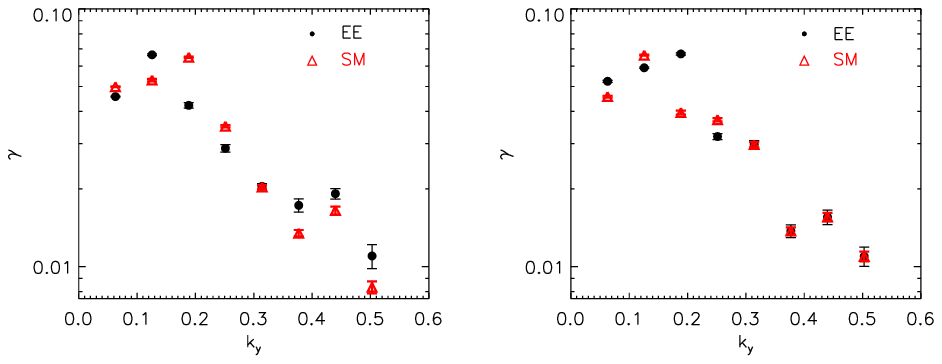}
\caption{Growth rate $\gamma$ of the first $8$ Fourier components of $E_{k_y}$, as a function of $k_y$ for the EE (black dots) and SM (red triangles) simulations. In the left (right) panel $E_{k_y}$ has been averaged over $x\in\left[0,L/2\right )$ ($x\in\left [L/2,L\right )$). The corresponding errorbars are also plotted.}
\label{fig:growthrate}
\end{figure*}
 
Growth rates $\gamma$ estimated by linearly fitting the quantity $E_{k_y}$ during the early exponential phase, are plotted in Fig.\ref{fig:growthrate}, for both EE (black dots) and SM (red triangles) simulations, as functions of $k_y$.
Left and right panels report the growth rates of the first eight Fourier components of the energy $E_{k_y}$ averaged over $x\in\left[0,L/2\right )$ and $x\in\left [L/2,L\right )$, respectively. As it can be appreciated from the two panels in Fig. \ref{fig:growthrate}, the development of the instability is not symmetric on the two shears: in particular, the fastest growing mode is not the same at the two shears, as first and second most unstable Fourier components are switched from left to right panel. Such asymmetry can be reasonably due to differences in the sign of ${\bm \omega} \cdot {\bf B}_0$ (the proton vorticity being ${\bm \omega}=\nabla \times {\bf u}$) at the two shears (positive in correspondence of the left shear and negative at the right one) \citep{Henri2013}. 
The fact that the growth rates are comparable in the two simulations is mainly due  to very similar velocity shear profiles adopted for the two runs.

\begin{figure*}[ht]
\centering
\includegraphics[width=\textwidth]{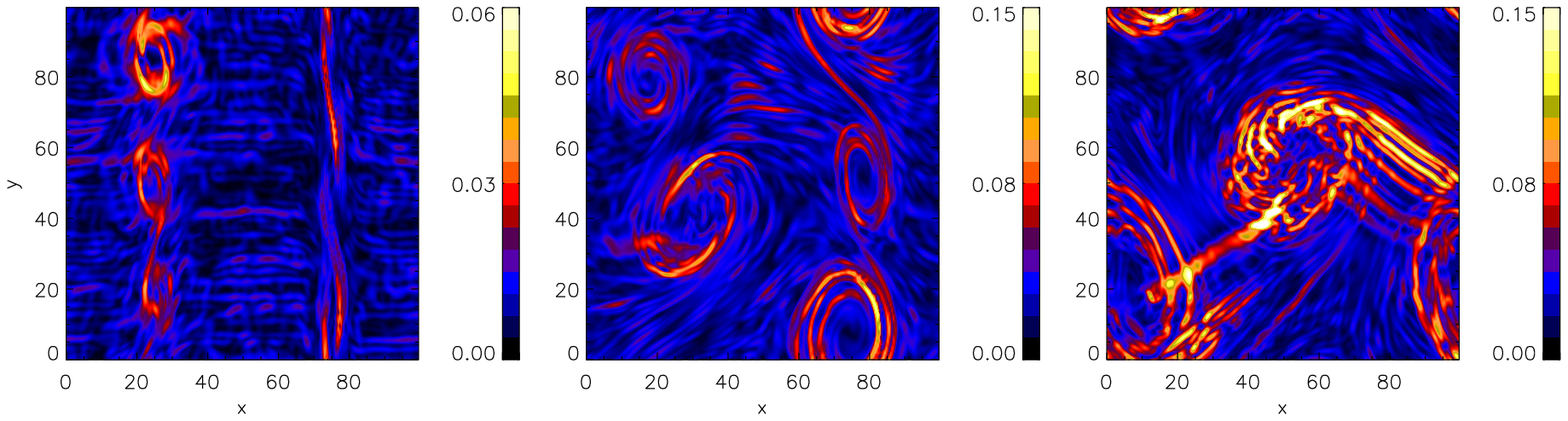}
\caption{Contour plot of $|{\bf j}|$ at three different times in the EE simulation. Left panel corresponds at time $t=80$, in the linear phase of the evolution of the instability, middle panel corresponds to $t=180$, when vortical structures start merging and, finally, right panel is at the end of the simulation, where vortices have collapsed in two large-scale structures and thin current filaments have been generated. }
\label{fig:contourJ}
\end{figure*}

\begin{figure}[ht]
\centering
\includegraphics[width=0.5\textwidth]{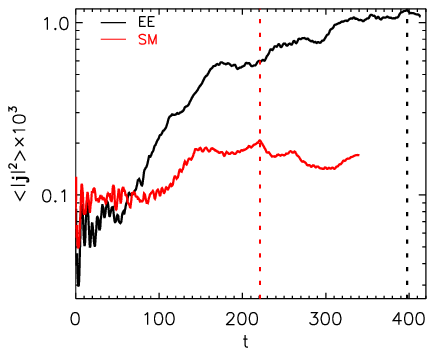}
\caption{Time evolution of the mean squared current density $\langle|{\bf j}|^2\rangle$ for the EE (black curve) and SM (red curve) simulation; vertical black (red) dashed line indicates the time at which the maximum level of turbulence is reached in the EE (SM) simulation. }
\label{fig:t-mj}
\end{figure}

\begin{figure}[ht]
\centering
\includegraphics[width=0.5\textwidth]{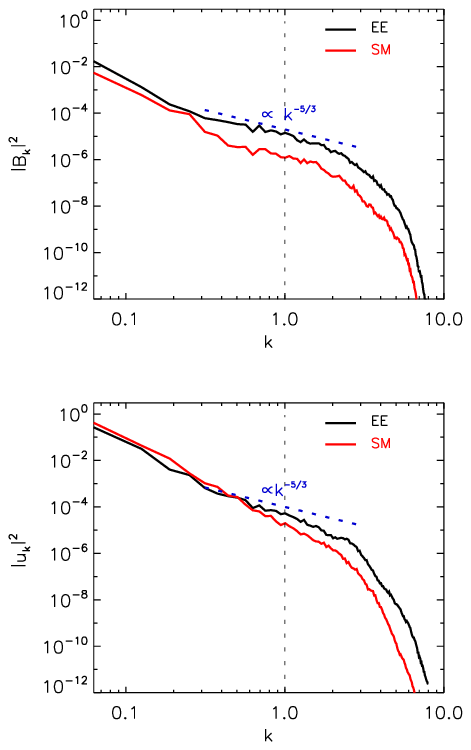}
\caption{Omnidirectional magnetic (top panel) and kinetic (bottom panel) energy spectra for the EE (black curve) and SM (red curve) simulation, taken at the time corresponding to the peak of $\langle|{\bf j}|^2\rangle$ (see Fig. \ref{fig:t-mj}). Kolmogorov expectation $k^{-5/3}$ is indicated in both panels by a blue-dashed line as a reference. In addition, the vertical black dashed line remarks the proton kinetic scale.}
\label{fig:spettri}
\end{figure}

 In the time evolution of the system, the initial exponential growth is followed by nonlinear saturation and later by a transition to turbulence. This can be appreciated in Fig. \ref{fig:contourJ}, which shows the contour plot of $|{\bf j}|$ for the EE simulation at three different times. The left panel of this figure corresponds to the time of the late linear phase of the instability and displays the formation of vortical structures in the shear regions (here the asymmetry between left and right shear is remarkable); in the middle panel, corresponding to the nonlinear saturation phase, vortices in both shears start merging and finally collapse in two distinct large-scale structures (right panel), in which short-scale filaments, whose size is few proton skin depth, are generated. 
In order to quantify the level of turbulence in the system, we looked at the time evolution of the mean squared current density $\langle |{\bf j}|^2\rangle$ ($\langle\cdots\rangle$ meaning spatial average). We first investigated separately the contribution of the two shears to $\langle |{\bf j}|^2\rangle$, noticing a very similar behaviour in the nonlinear and late time stage of the simulation. Thus, we decided to average over the whole spatial domain. In Fig. \ref{fig:t-mj}, we report the time evolution of $\langle |{\bf j}|^2\rangle$ for the EE (black curve) and SM (red curve) simulations.
Here, significant differences are recovered between EE and SM cases: in fact, generation of turbulence seems to be inhibited in the case of the SM initial condition, for which the saturation value of $\langle|{\bf j}|^2\rangle$ is about one order of magnitude lower than in the EE case. To better point out this effect, in Fig. \ref{fig:spettri} we show the omni-directional spectra of magnetic (top panel) and kinetic (bottom panel) energy, evaluated at the time in correspondence of the vertical black (red)-dashed line in Fig. \ref{fig:t-mj} for the EE (SM) simulation. These spectra (Kolmogorov expectation $k^{-5/3}$ is indicated by a blue dashed line as a reference) clearly show a larger energy content (about an order of magnitude in the inertial range) for the EE case (black lines) as compared to the SM case (red lines). Moreover, although in both EE and SM cases the spectral energy is peaked at low wavenumbers, a Kolmogorov-like spectrum is observed for about a wavenumber decade. 

The inhibition of turbulence generation occurring in the simulation with the SM initial condition may be due to the fact that $f_0^{^{(SM)}}$ is not an exact equilibrium DF for the HVM equations in presence of a velocity shear. Indeed, as discussed in previous works (see, for example, \citet{Cerri2013} and \citet{Malara2018}), this feature naturally induces oscillations on time scales of the order of the $\Omega_{cp}^{-1}$ and on spatial scales of the order of the proton skin depth in the proton density, bulk speed, and also higher order moments. These oscillations may lock the energy at particular wavenumbers, preventing it from efficiently contributing to the turbulent cascade.
As the HVM code retains kinetic effects on protons, the question arises whether the development of turbulence across $d_p$ produces deformations of the proton DF. In the following, we seek for local deviations from Maxwellianity and for the generation of sharp gradients in the proton velocity distribution. We recall that, at $t=0$, $f_0^{^{(EE)}}$ departs from a Maxwellian in the shear regions, while the perturbed SM initial condition, being setup as a sum of two distinct Maxwellians, is not a Maxwellian. Then, we investigate: (i) if distortions from the Maxwellian shape increase as turbulence develops, and (ii) if these distortions remain confined in the shear regions. In order to quantify deviations from a Maxwellian, we employ the non-Maxwellianity indicator introduced in \citet{Greco2012} and defined as:
\begin{equation}
    \epsilon(x,y,t)=\frac1 n\sqrt{\int (f-g)^2 d^3v}
\end{equation}
where $g$ is the Maxwellian DF associated with $f$, i.e.,  which has the same velocity moments (density, bulk velocity and temperature) as $f$.  

\begin{figure*}[ht]
\centering
\begin{minipage}[hb]{0.45\linewidth}
   \centering
\includegraphics[width=\textwidth]{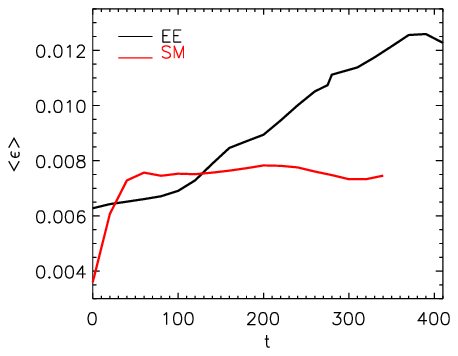}
\end{minipage}
\begin{minipage}[hb]{0.45\linewidth}
   \centering
\includegraphics[width=\textwidth]{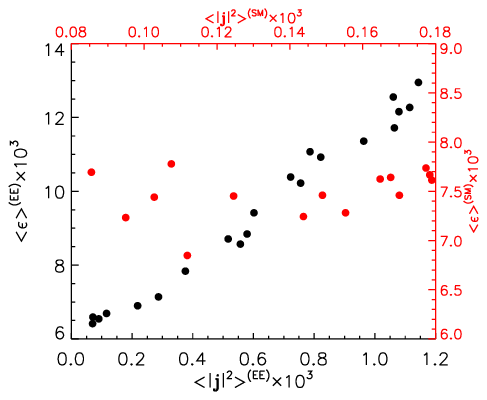}
\end{minipage}
\caption{Left: time evolution of the non-Maxwellian indicator averaged over the whole spatial domain $\langle\epsilon\rangle$ for the EE (black curve) and the SM (red curve) simulation. Right: scatter plot of $\langle\epsilon\rangle$ as a function of $\langle|{\bf j}|^2\rangle$ for the EE  (black dots) and SM (red dots) simulation. }
\label{fig:eps}
\end{figure*}

In the left panel of Fig. \ref{fig:eps} we show the time evolution of $\langle\epsilon\rangle$ ($\langle\cdots\rangle$ meaning average over the whole spatial domain $D$), for both EE and SM simulations. At $t=0$ $\langle\epsilon\rangle$ starts from a non-zero value for both EE and SM simulations, as anticipated above. Both quantities then grow in time, indicating efficient generation of non-Maxwellian features during the EE simulation, while saturating after the initial growth in the case of the SM simulation. 
However, the saturation level of $\langle \epsilon \rangle$ is larger for the EE case with respect to the SM one, this suggesting that the generation of non-Maxwellian features in the DF is much more efficient in the former case.
In the right panel of the same figure, we present the scatter plot of $\langle\epsilon\rangle$ versus $\langle|{\bf j}|^2\rangle$, showing that the increase of the non-Maxwellianity indicator appears to be well correlated in time with the increase of the level of turbulence in the system (time Pearson correlation coefficient is $C_t\simeq 0.99$) for the EE simulation (black dots). On the other hand, for the SM case (red dots) the correlation between the two quantities is not as high as in the previous case ($C_t\simeq 0.48$), resulting in an almost flat trend in the figure. This last evidence provides clear indication that, as turbulence brings energy towards small wavelengths, the proton DF departs more and more from local thermodynamic equilibrium, in the case of the EE simulation.

\begin{figure*}[ht]
\centering
\includegraphics[width=\textwidth]{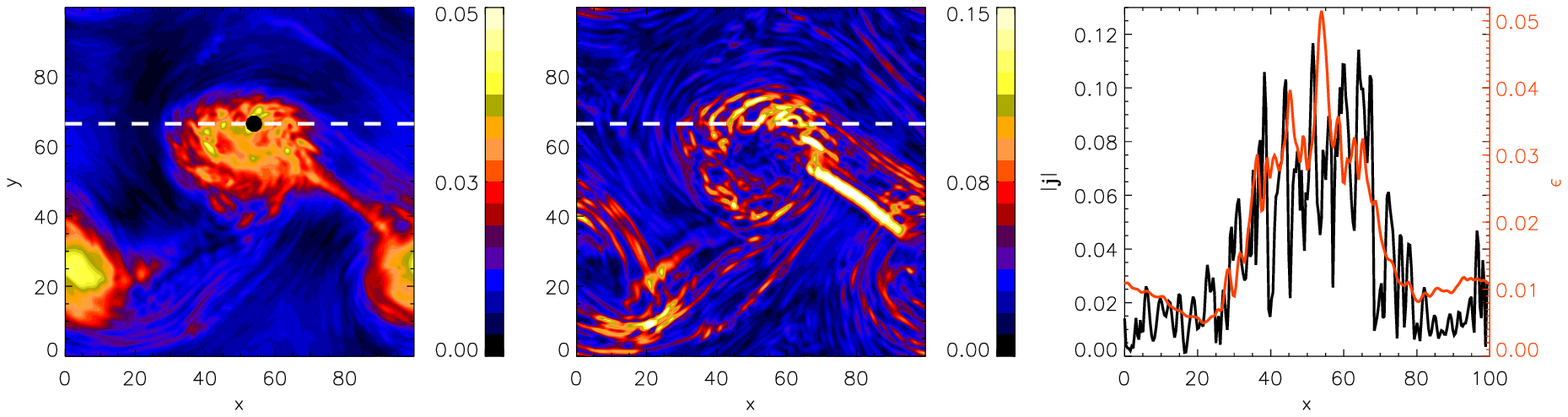}
\caption{Contour plot of $\epsilon$ (left panel) and of $|{\bf j}|$ (middle panel) at the time of the maximum level of turbulence in the EE simulation. In the right panel, cuts of $|{\bf j}|$ (black curve) and of $\epsilon$ (red curve), taken along the horizontal white-dashed paths in left and middle panels.}
\label{fig:contour_eps-J}
\end{figure*}

From now on, we will limit our discussion to the case of the EE simulation and investigate in more detail the role of kinetic effects in shaping the proton velocity distribution. We then looked at the spatial patterns of $\epsilon$ and $|{\bf j}|$ at a fixed instant of time. In Fig. \ref{fig:contour_eps-J}, we report the contour plot of $\epsilon$ (left panel) and $|{\bf j}|$ (middle panel) at the time of the maximum level of turbulence in the system (vertical black-dashed line in Fig. \ref{fig:t-mj}). The spatial features of the two quantities are very similar, with peaks of $\epsilon$ concentrated inside the vortical structures of $|{\bf j}|$; the horizontal cuts (right panel) of $\epsilon$ (red) and $|{\bf j}|$ (black), taken along the horizontal white-dashed lines in the left and middle panels of this figure, confirm that the peaks in the non-Maxwellianity indicator thicken inside the vortical current structures. 

\begin{figure*}[hb]
\centering
\includegraphics[width=\textwidth]{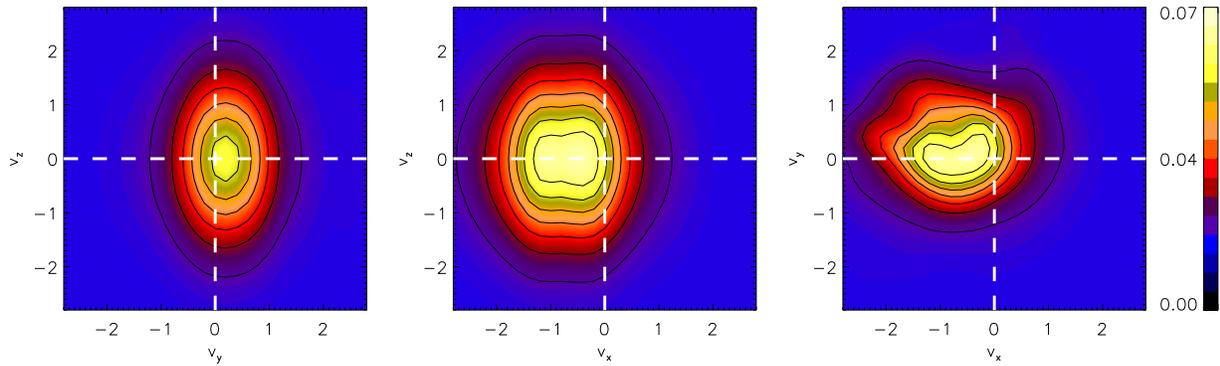}
\caption{Two-dimensional contour plots of the proton DF for the EE simulation at the time of the maximum level of turbulence (black dashed line in Fig. \ref{fig:t-mj} ) and at the spatial point of the maximum of $\epsilon$ (black dot in Fig. \ref{fig:contour_eps-J}). $(v_y,v_z)$ plane is reported in the left panel, $(v_x,v_z)$ plane in the middle panel and $(v_x,v_y)$ plane in the right panel. }
\label{fig:2DVDF}
\end{figure*}

\begin{figure*}[ht]
\centering
\includegraphics[width=0.7\textwidth]{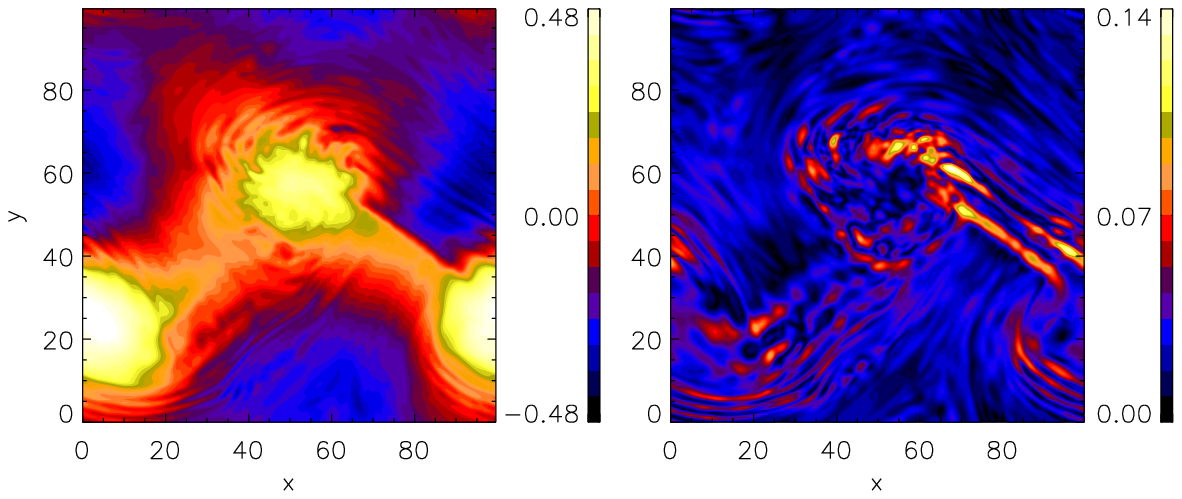}
\caption{Two-dimensional contour plots of the temperature anisotropy index $A$ (left) and agyrotropy parameter $\sqrt{Q}$ (right) evaluated at the time of maximum activity for the EE simulation. }
\label{fig:contour_anys-agyr}
\end{figure*}

In Fig. \ref{fig:2DVDF} we show how the proton DF looks like at the time of the maximum level of turbulence in the system and at the spatial point where $\epsilon$ is maximum (black dot in the left panel of Fig. \ref{fig:contour_eps-J}). Here, the 2D contour plots of $f$ are shown in the $(v_y,v_z)$ plane for $v_x=0$ (left panel), in the $(v_x,v_z)$ plane for $v_y=0$ (middle panel) and in the $(v_x,v_y)$ plane for $v_z=0$ (right panel). Left and middle panels of this figure display generation of significant temperature anisotropy, 
while the right panel shows peculiar deformations, with the generation of modulations and sharp velocity gradients, driven by the interaction of particles with the field fluctuations.

A deep analysis of both anisotropy and agyrotropy of the proton DF allows to provide quantitative information on the features observed in the contour plots of Fig. \ref{fig:2DVDF}.  In the left panel of Fig. \ref{fig:contour_anys-agyr}, we plot the spatial variation of the anisotropy index $A(x,y)=1-T{_\perp}/T_{\parallel}$, where $T_\perp$ and $T_\parallel$ are the temperatures in the direction transverse and parallel to the local magnetic field, respectively. This plot is calculated at the time corresponding to the maximum level of turbulence in the system. Negative (positive) values of $A$ correspond to $T_\perp > T_\parallel$ ($T_\perp < T_\parallel$). 
In the right panel of the same figure, we show the agyrotropy parameter $\sqrt{Q}$, linked to the off-diagonal terms of the pressure tensor , where $Q$ is defined as:
\begin{equation}
    Q=\frac{P_{xy}^2+P_{xz}^2+P_{yz}^2}{P_\perp^2+2P_\perp P_\parallel};
    \label{eq:gyrotropy}
\end{equation}
here, $P_{ij}$ are the components of the pressure tensor in the reference frame in which one of the axes is along the local magnetic field \citep[see][for more details]{Swisdak2016}. The agyrotropy parameter ranges from $0$ to $1$, where $\sqrt{Q}=0$ and $\sqrt{Q}=1$ correspond to fully gyrotropic configurations and maximum agyrotropy, respectively. 
It can be easily noticed that iso-contours of $A$ and $\sqrt{Q}$ exhibit a pattern similar to those visible in the contour plots of $\epsilon$ and $|{\bf j}|$ (left and middle panels in Fig. \ref{fig:contour_eps-J}), respectively. Indeed, $A$ reaches its highest values in the center of each vortex (same as $\epsilon$), while $\sqrt{Q}$ achieves its highest value at the edges of the vortices (similar to $|{\bf j}|$). Moreover, in correspondence of the maximum value of $\epsilon$, $\sqrt{Q}$ displays highly non gyrotropic features of the proton DF and $A>0$ suggests a significant elongation of the DF in the direction parallel to the local magnetic field.
Incidentally, we notice that within the shear layers the EE DF is also anisotropic, but with $T_\perp > T_{||}$ \citep{Malara2018}.

\begin{figure*}[ht]
\begin{minipage}[ht]{0.32\textwidth}
\centering
  \includegraphics[width=0.5\textwidth]{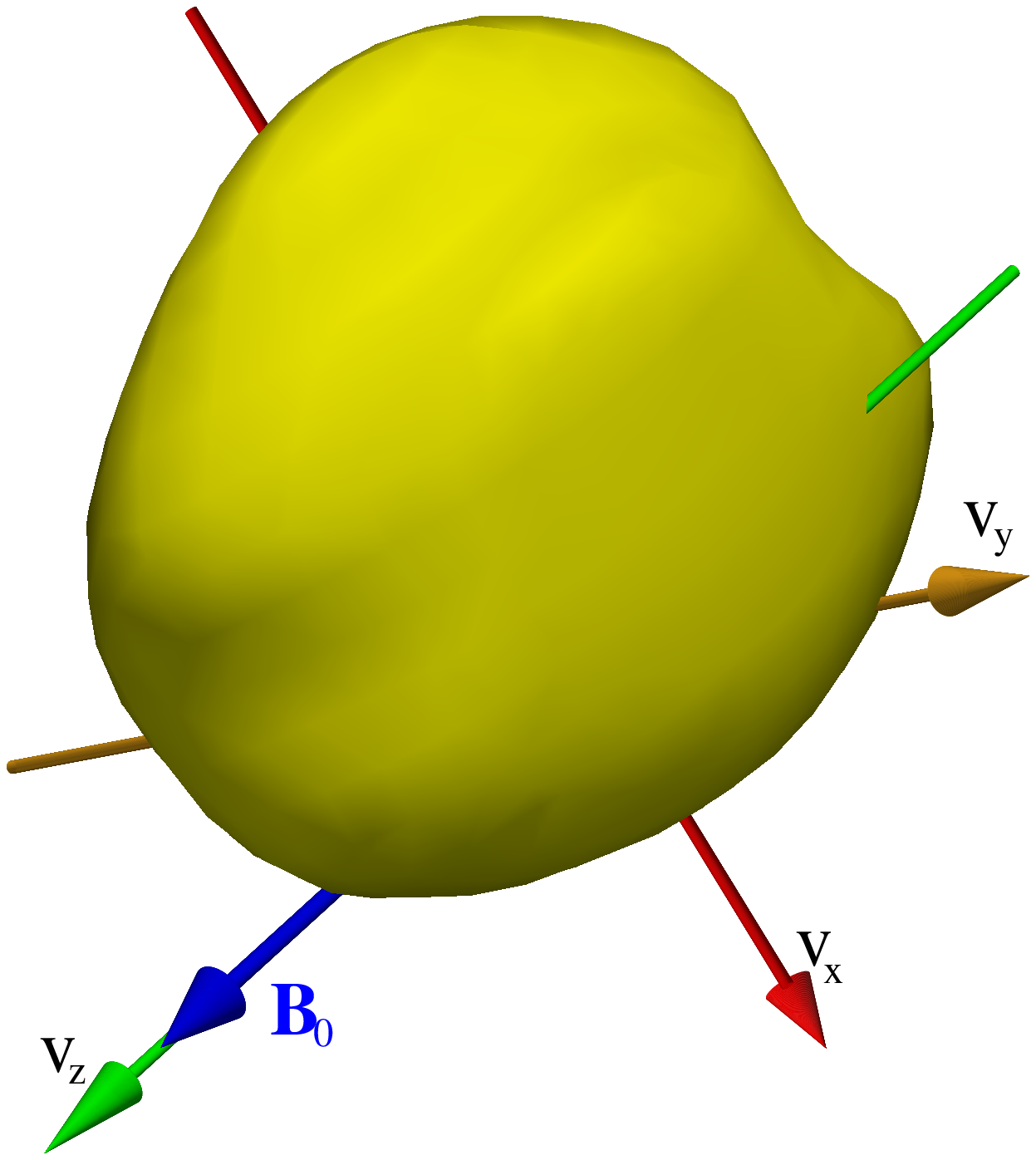}
\end{minipage}
\begin{minipage}[ht]{0.32\textwidth}
   \centering
   \includegraphics[width=0.5\textwidth]{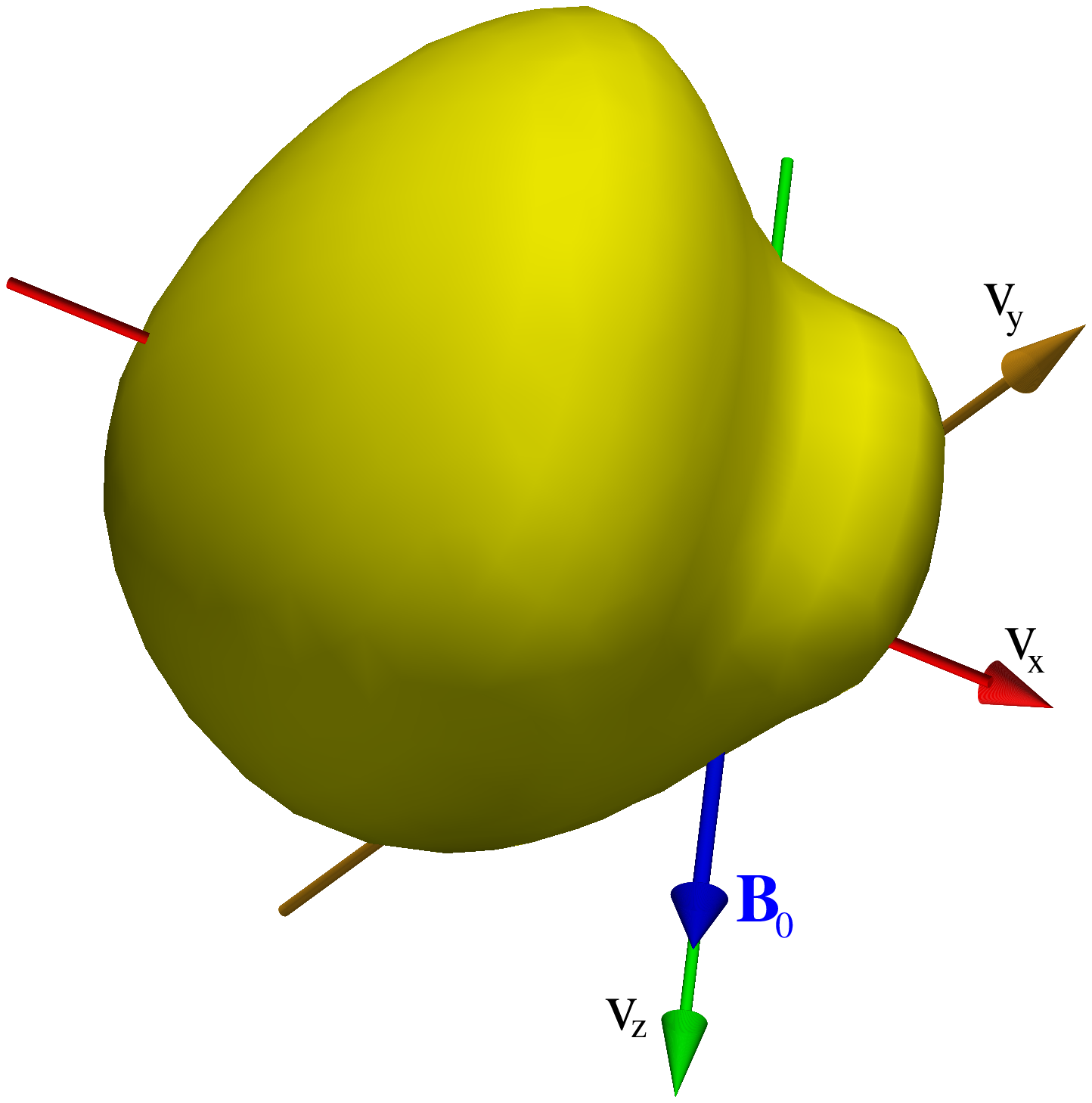}
\end{minipage} 
\begin{minipage}[ht]{0.32\textwidth}
   \centering
   \includegraphics[width=0.5\textwidth]{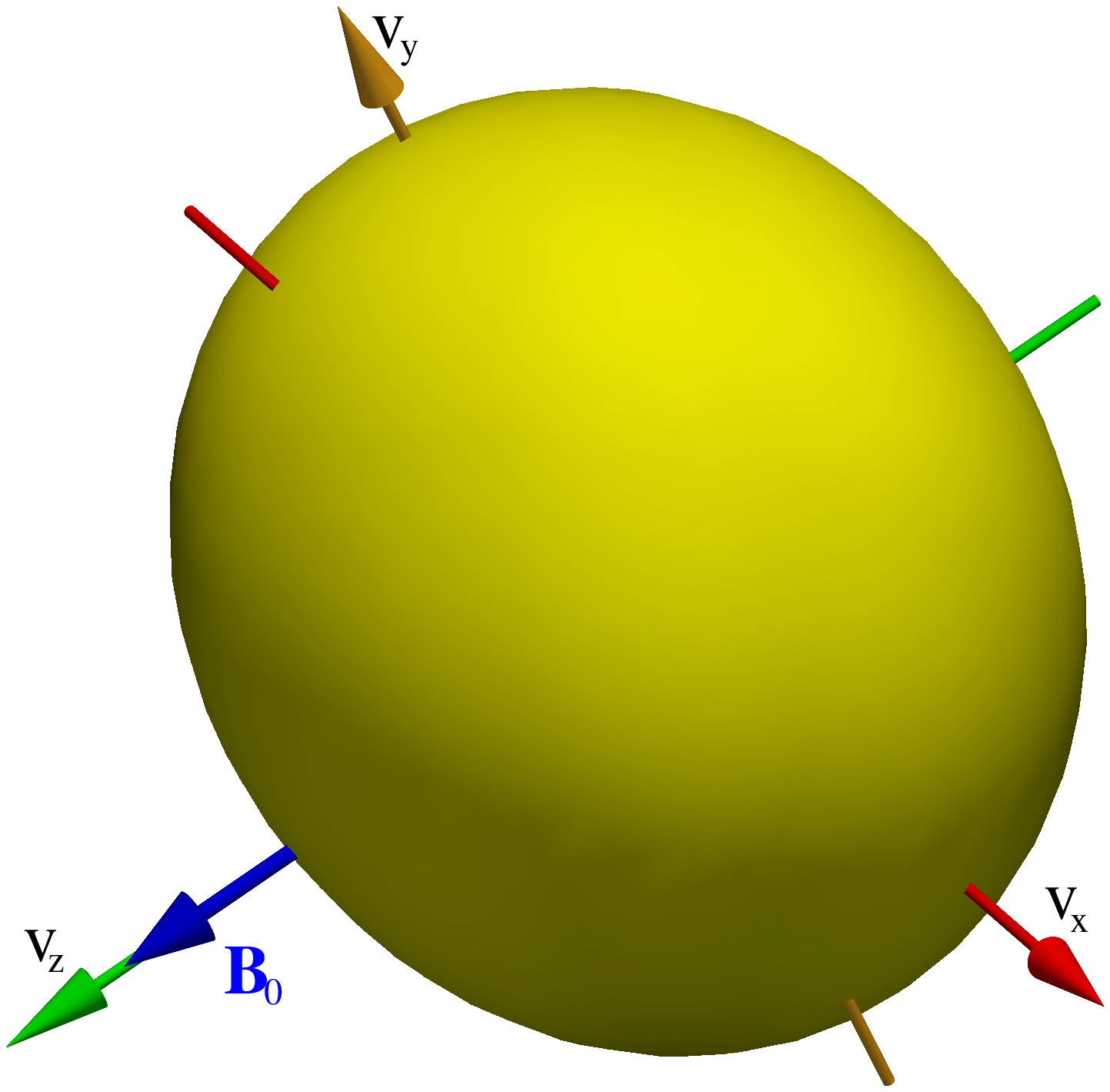} 
 \end{minipage}
\caption{Iso-surface plots of three proton DFs for the EE simulation at the same time of Fig. \ref{fig:2DVDF}. From left to right the DFs have been evaluated at the spatial points corresponding to $A>0$, $A=0$, $A<0$, respectively. The blue arrow indicates the direction of the background magnetic field.}
\label{fig:fd3D}
\end{figure*}

Finally, it is interesting to look at the shape of the velocity DF in correspondence of three different values of $A$, corresponding to $A>0$, $A=0$ and $A<0$. In Fig. \ref{fig:fd3D}, we report the three-dimensional velocity iso-surface of the proton DF at spatial point $(x,y)=(53.9,66.4)$ for $A>0$ (left panel), $(x,y)=(53.9,73.04)$ for $A=0$ (middle panel) and at the coordinate $(x,y)=(53.9,3.91)$ for $A<0$ (right panel). These plots show that departures from Maxwellianity are not only simply related to temperature anisotropy, but the DF displays a highly irregular shape. This is particularly visible in the case $A=0$ (middle panel), where complex structures in the velocity space are visible, in spite of the temperature isotropy.

\section{Summary and conclusions} \label{sec:concl}
In this paper we have studied the nonlinear and turbulent stage of the KHI and the related kinetic effects produced on particles, by means of Hybrid Vlasov-Maxwell simulations at proton scales. In particular, we have considered an unperturbed configuration where the background magnetic field is perpendicular to the shear flow. This configuration is Kelvin-Helmholtz unstable regardless of the value of velocity jump across the shear layer, at least in the MHD case.

In kinetic descriptions of KHI, the unperturbed configuration has been often represented by means of a SM DF, though this is not a stationary solution. We have shown that, when an exact equilibrium solution is chosen to initialize the system, relevant effects on the dynamics of KHI appear. To highlight this point, we have compared KHI simulations with two different onsets for the DF, namely (i) the EE solution and (ii) a SM DF, which is not a stationary solution. Due to spurious fluctuations, the non-stationary solution tends to inhibit turbulence that develops during the nonlinear phase of KHI. The enhancement of turbulent activity in the EE simulation can be observed in the spectra, where the magnetic and kinetic energy in the inertial range of the EE simulation are roughly one order of magnitude larger than in the SM case. 
Moreover, considering the mean square current density, which is mainly determined by small scales, we found that $\langle |{\bf j}|^2\rangle$ reaches a much higher level in the EE simulation when compared with the SM case. This is a further indication of an enhanced turbulent activity in the EE case.

Differences between the two cases have also been found during the linear stage of KHI. Growth rates of unstable modes have different values in the EE and SM cases, also according to the relative vorticity-magnetic field orientation. In particular, in the shear layer where ${\bm \omega}$ is parallel to ${\bf B}$  the most unstable mode has wavelength $\lambda_y=L/2$ in the EE case and $\lambda_y=L/3$ in the SM case, while the reverse holds in the shear layer where  ${\bm \omega}$ is antiparallel to ${\bf B}$.

As a consequence of the efficient energy transfer towards shorter scales in the EE simulation, the proton velocity distribution significantly departs from the local thermodynamic equilibrium. In particular, the enhancement of the turbulent activity leads to stronger deviations from the Maxwellian configuration, mainly located near regions of high magnetic stress, i.e. strong current sheets, and not in correspondence of the shears (as one would expect). A similar behavior has been observed in recent space observations by MMS data in both KHI in the Earth’s magnetosphere \citep{Sorriso2019} and in the turbulent dynamics of the Earth’s magnetosheath \citep{Perri2020}. 

A detailed analysis of the proton DF has shown the presence of significant temperature anisotropies and agyrotropies. We have observed that the DFs display strong deformations where $T_{\parallel}$ is higher than or close to $T_{\perp}$, these corresponding to spatial position where $\epsilon$ reaches large values. In the other regions (where $T_{\perp}>T_{\parallel}$), instead, the non-maxwellianity parameter is close to zero, thus indicating a slightly distorted distribution. Indeed, there, the VDF is quite smooth and it only shows an elongation in the perpendicular direction with respect to the local magnetic field. At such points, the field seems to not play a significant role in shaping the proton DF.

Our numerical results contribute to give a better understanding of the proton kinetic dynamics and energy transfer mechanisms towards small scales, with a main focus on the turbulence enhancement due to the instability, observed when the exact hybrid-Vlasov equilibrium is used as the initial unperturbed state for the simulation.
These results represent a first step towards a comparison between synthetic and MMS \textit{in situ} data. Owing to the shear merging observed at late times, our present results cannot be used to make a point-to-point comparison with MMS observations to study the nonlinear effects of the instability. Indeed, in the Earth's environment the regions of vorticity parallel and antiparallel with respect to the magnetic field are located at the two flanks of the magnetopause. 
In a future work, we plan to use a larger box domain in order to keep the shear layers well separated for the entire simulation time and compare in detail our numerical results with a KH event observed by the MMS spacecrafts, as in \citet{Henri2013}.

Moreover, further studies concerning the onset of KHI in a collisionless plasma will be conducted in a fully 3D physical space, i.e. in the full 6D phase space, where we expect a much more complex and rich dynamics.

\section*{Acknowledgments} Numerical simulations have been run on Marconi supercomputer at CINECA (Italy) within the ISCRA projects: IsC68\_TURB-KHI and IsB19\_6DVLAIDA. This work has received funding from the European Unions Horizon 2020 research and innovation programme under grant agreement no. 776262 (AIDA, www.aida-space.eu)

\end{document}